\documentclass[reprint,amsmath,amssymb,aps,superscriptaddress,pre]{revtex4-2}

\usepackage{graphicx}
\usepackage{xcolor}
\usepackage{dcolumn}
\usepackage{bm}
\usepackage{booktabs}
\usepackage{sidecap}
\sidecaptionvpos{figure}{t}
\usepackage[pdftex,bookmarks,colorlinks,breaklinks]{hyperref} 
\hypersetup{linkcolor=red,citecolor=blue,filecolor=dullmagenta,urlcolor=blue} 
\hypersetup{pdftoolbar=true,pdfpagemode=UseNone,pdfstartview=FitH, colorlinks=true,extension=}

\begin{document}

\title[Synchronization driven acoustics]{Synchronization driven acoustics: \texorpdfstring{\\}{} The nonlinear scattering of a self-oscillating meta-atom}
\author{Alexander K. Stoychev}
\email[]{astoychev@ethz.ch}
\affiliation{CAPS Laboratory, Department of Mechanical and Process Engineering, ETH Zürich, 8092 Zürich, Switzerland}

\author{Xinxin Guo}
\affiliation{CAPS Laboratory, Department of Mechanical and Process Engineering, ETH Zürich, 8092 Zürich, Switzerland}

\author{Ulrich Kuhl}
\affiliation{CAPS Laboratory, Department of Mechanical and Process Engineering, ETH Zürich, 8092 Zürich, Switzerland}
\affiliation{Université Côte d’Azur, CNRS, Institut de Physique de Nice (INPHYNI), 06200, Nice, France}

\author{Nicolas Noiray}
\email[]{noirayn@ethz.ch}
\affiliation{CAPS Laboratory, Department of Mechanical and Process Engineering, ETH Zürich, 8092 Zürich, Switzerland}

\date{\today}
\begin{abstract}
In this study we demonstrate a self-oscillating acoustic meta-atom functioning as an amplifying transistor, where a steady external flow serves as a control signal to switch between reflective (off-state) and transmissive (on-state) regimes. In the on-state, an acoustic limit cycle synchronizes with incident sound waves. This process governs the energy transfer across the device, with a transmission bandwidth dictated by the synchronization region in parameter space (Arnold tongue). Our experimental measurements reveal nonlinear dependence on the incident wave amplitude, enabling perturbation filtering therein and stabilizing downstream acoustic power. All experimentally observed phenomena are quantitatively described by a nonlinear Liénard-type oscillator featuring saturable gain and linear loss, where the essential parameters can be estimated by independent measurements. This work may offer a paradigm shift in acoustic metamaterials research by leveraging self-oscillation and synchronization processes. Bridging those key concepts from nonlinear dynamics and complex systems with active metamaterial design in acoustics and related disciplines, may establish a broadly applicable framework of field-independent mechanisms for wave manipulation.
\end{abstract}
\maketitle

\section{Introduction}\label{sec:Introduction}
Acoustic metamaterials have been a source of fruitful observations over the past decade \cite{Cummer2016,Ma2016,Assouar2018,Nassar2020,Xue2022}, with some of the utilized concepts originating in, or pertinent to electromagnetics \cite{Zheludev2012,Kadic2019}. Negative refraction is one example of a breakthrough transcending its original domain; initially proposed \cite{Pendry2000} and implemented \cite{Fang2005} for superlensing in the (near) visible spectrum, with subsequent applications in acoustics \cite{Fang2006,Liu2000} and elastic wave propagation \cite{Lakes2001}. Another case in point could be found in bias driven reciprocity breaking, which has been demonstrated for both electromagnetic \cite{Sounas2013} and acoustic scattering \cite{Fleury2014}. Moreover, a recent study introducing a phononic topological transistor \cite{Pirie2022} has developed a novel methodology, based on Dirac metamaterials with accidental degeneracy, which is claimed to be directly transferable to photonic crystals and electronic logic.

Although employing nonlinear effects for wave propagation control is a standard practice, e.g., to break reciprocity \cite{Liang2009,Liang2010,Popa2014}, synchronization based developments are not widely adopted in acoustic metamaterials \cite{Assouar2018,Nassar2020,Xue2022} despite their ubiquity \cite{Strogatz1993,Collins1995,Singer1999,Strogatz2000,Strogatz2001,Pikovsky2001}, however, there are recent results in other disciplines \cite{Zheng2023,Salcedo2025}. While saturable gain has been utilized for $\mathcal{PT}$-symmetric scattering, demonstrating, e.g., unidirectional invisibility \cite{Zhu2014,Auregan2017} and non-invasive sensing \cite{Fleury2015}, the use of self-oscillators in acoustic metadevices has been limited. However, interactions with self-sustained limit cycles offer a unique venue for novel (and domain independent) wave phenomena, e.g., illustrating specific scattering states \cite{Harrison1990,Uzunov2014}, robust wireless power transfer \cite{Assawaworrarit2017}, nonlinear degeneracies \cite{Uzdin2012,Wang2019,Bai2023,Bai2024} or manipulation of objects~\cite{Bode2011,Cui2024}.

Indeed, synchronization driven acoustic scattering has already been shown in the context of band limited cloaking \cite{Pedergnana2025}, loss-compensated reciprocity breaking \cite{Pedergnana2024} and superradiance \cite{Pedergnana2023}. While those goals can be achieved by alternative means, e.g. \cite{Cromb2020}, exploiting the nonlinear dynamics of limit cycle oscillators offers an important paradigm shift promising to reconcile the rich landscape of synchronization in complex systems with metamaterials research.

\begin{figure}[b]
    \includegraphics[width=\linewidth]{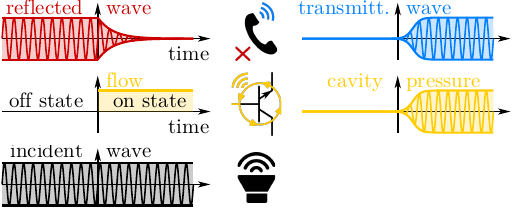}
    \caption{Scattering behavior of the nonlinear meta-atom. In the on-state, transmission and reflection are governed by the synchronization between the self-oscillator and incident sound waves. In the off-state, the weak coupling between the meta-atom's (passive) resonance and the external acoustic field leads to high reflection and negligible transmission. Note that in the on-state, as the incident amplitude approaches zero, the synchronization band shrinks around the autonomous self-oscillation frequency and the scattering coefficients exhibit singularity due to the finite amplitude radiation.}
    \label{fig:Overview}
\end{figure}

In the following we investigate the nonlinear scattering properties of a self-oscillating meta-atom, showing that such a device can be utilized as an (amplifying) acoustic transistor. In particular, a constant control signal in the form of a steady external flow [analogous to direct current (DC)], can be employed as a means of modifying the system's response from fully reflective to fully transmissive [Fig.~\ref{fig:Overview}]. The bandwidth of this effect is determined by the synchronization region (Arnold tongue) in parameter space [cf. Apps.~\ref{app:ModelCoefficients}--\ref{app:SynchronizationSpectrum}], which can be influenced by the design of the meta-atom and it's coupling interface to the waveguide. Furthermore, the scatterer exhibits nonlinear dependence on the incident wave amplitude, filtering acoustic power perturbations and stabilizing the input signal for devices downstream thereof.

The aims of this study go beyond a demonstration of an acoustic transistor; this has already been achieved by a combination of nonlinearity and filtering (for monochromatic signals) \cite{Liang2014}, controlled topological transitions \cite{Pirie2022}, and nonreciprocal Willis coupling \cite{Wen2023} (diode). Our main focus is to illustrate the untapped potential of synchronization driven interactions between self-oscillators and external fields supporting wave propagation, and to introduce a theoretical framework capturing this dynamical phenomenon in fine quantitative detail.

The synchronization process of the utilized self-oscillating meta-atom shares some similarities with injection-locking of microwave oscillators, i.e., a free-running RF oscillator pumped by DC bias (e.g. Gunn diode) that locks in phase and frequency to an injected signal \cite{Perlman1970,Kurokawa1973}. Just as Gunn diodes can be guided by an external RF signal, our acoustic self-oscillator can lock to an external sound field. Specifically, the acoustic limit cycle (pumped by steady air flow) is analogous to the RF oscillator, and the incident wave plays the role of the injection signal. When synchronization occurs within the corresponding parameter space region (Arnold tongue), the phase/frequency is pulled to match the input, mirroring the behavior of various known systems \cite{Li2022,Navarro2022,Euler2024}. The aforementioned highlights this type of system as a fluid-mechanical analogue of a DC-to-AC oscillator subject to injection locking.
\section{Results}\label{sec:Results}
\begin{figure}
    \includegraphics[width=\linewidth]{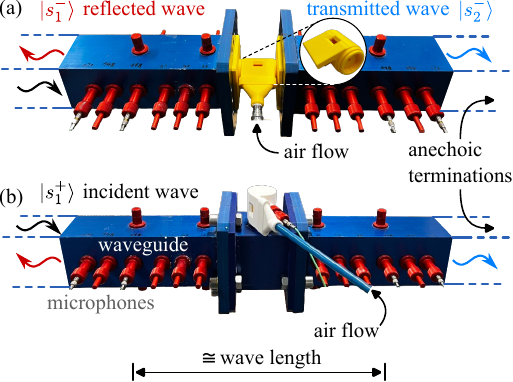}
    \caption{Depiction of the experimental setup utilized to determine the scattering properties of the self-excited meta-atom in the (a) direct and (b) side-branch configuration.}
    \label{fig:ExperimentalSetup}
\end{figure}

To illustrate the concept intuitively, we implement the self-oscillating meta-atom as a whistle inside a scattering arrangement [Fig.~\ref{fig:ExperimentalSetup}]. For details of the whistle itself see App.~\ref{app:WhistleGeometry}. Specifically, in the on-state, a steady external flow [the control signal] establishes a hydrodynamic shear layer over the aperture on top of the quasi rigid cavity, which in turn serves as a mechanism for partially converting the steady kinetic energy of the flow into an (oscillating) acoustic source [cf.\ \cite{Bourquard2021,Aperture}]. This process results in a self-sustained limit cycle of uniform cavity pressure, which is acoustically coupled to the adjacent waveguide via two ports. This coupling allows interactions with waves propagating in the surrounding medium, leading to (i) lossy, (ii) unitary or (iii) superradiant scattering states. Conversely, in the off-state, the weak wave-resonance coupling via the ports and the large impedance mismatch between the waveguide and the whistle's cavity result in negligible energy exchange, inhibiting the transfer of acoustic power across the meta-atom.

The setup for studying the scattering properties of our meta-atom is depicted in Fig.~\ref{fig:ExperimentalSetup}, and consists of two channels with a square cross section of 62 mm width which are terminated by anechoic ends (not shown), simulating semi-infinite waveguides \cite{Bourquard2021}. The waveguide supports only one propagating mode in the frequency range of interest. The self-oscillating meta-atom (whistle) is supplied with air via a volume flow controller [Bronkhorst Flexi-Flow] and is coupled to the channel through 4\,mm [Fig.~\ref{fig:ExperimentalSetup}(a)] or 7\,mm [Fig.~\ref{fig:ExperimentalSetup}(b)] orifices on the sides, respectively. The system is subject to an incident acoustic plane wave generated by a compression driver [CP850Nd, Beyma] located inside the anechoic termination. The temporal evolution of the resulting acoustic pressure is recorded, at a sampling rate of 50\,kHz, with three microphones [46BD-FV G.R.A.S.] in each waveguide section, and the Riemann invariants of the incident and scattered acoustic waves are determined via spectral decomposition at the forcing frequency. The (uniform) pressure inside the cavity is measured as well.
\begin{figure*}
    \includegraphics[width=\linewidth]{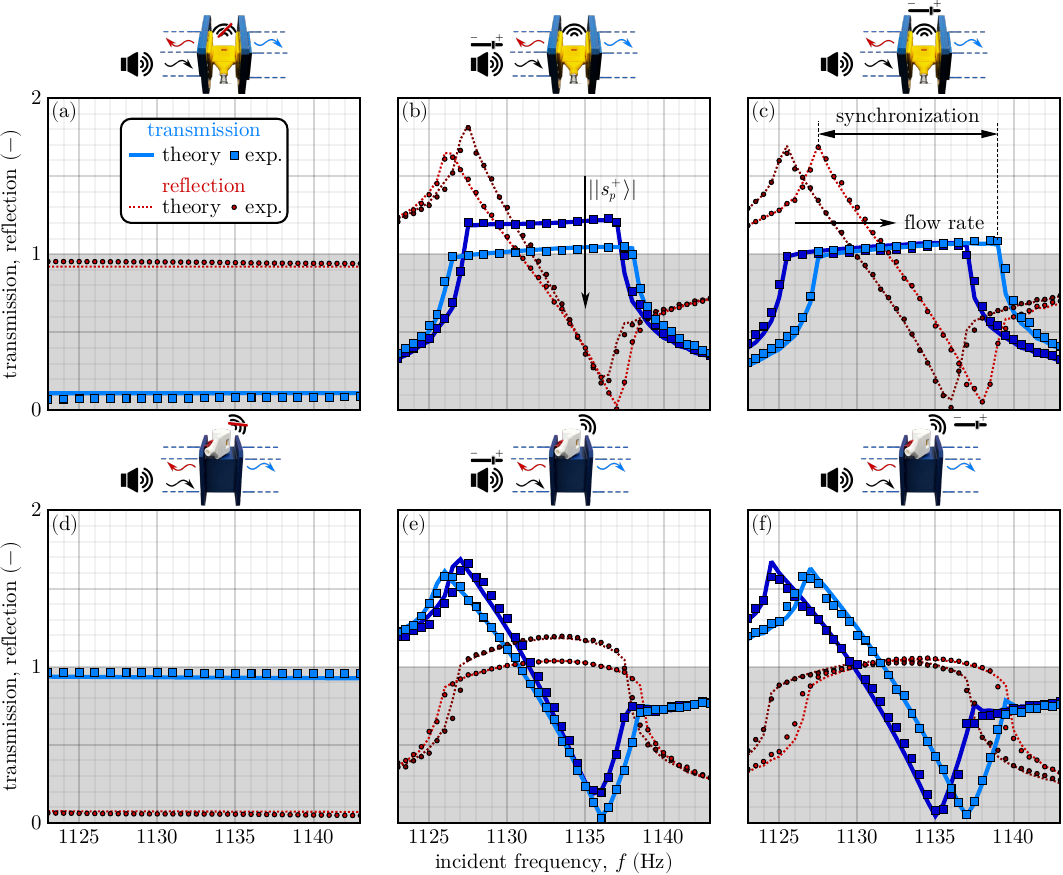}
    \caption{The top panels depict the scattering coefficients of the direct configuration [Fig.~\ref{fig:ExperimentalSetup}(a)] for (a) vanishing control signal [off-state; no volume flow], (b) nominal control signal $[\dot{Q}=9~\mathrm{l/min}]$ and varying excitation amplitude $\left[\left\vert\left\vert s_p^{\mathrm{+}}\right\rangle\right\vert\approx4~\mathrm{Pa};5~\mathrm{Pa}\right]$, and (c) varying control signal $[\dot{Q}=8.94~\mathrm{l/min};9.02~\mathrm{l/min}]$ and nominal excitation amplitude $\left[\left\vert\left\vert s_p^{\mathrm{+}}\right\rangle\right\vert\approx5~\mathrm{Pa}\right]$. Analogously, the bottom panels show the scattering coefficients of the side-branch configuration [Fig.~\ref{fig:ExperimentalSetup}(b)] for (d) vanishing control signal, (e) nominal control signal $[\dot{Q}=10~\mathrm{l/min}]$ and varying excitation amplitude $\left[\left\vert\left\vert s_p^{\mathrm{+}}\right\rangle\right\vert\approx6.5~\mathrm{Pa};7.5~\mathrm{Pa}\right]$, and (f) varying control signal $[\dot{Q}=9.88~\mathrm{l/min};9.94~\mathrm{l/min}]$ and nominal excitation amplitude $\left[\left\vert\left\vert s_p^{\mathrm{+}}\right\rangle\right\vert\approx7.5~\mathrm{Pa}\right]$. Shades of blue represent transmission, with square markers depicting measurement points and solid lines - theoretical calculations. Conversely, shades of red denote reflection, with circular markers being measurement points and solid lines - solutions of Eqs.~\eqref{eq:LimitCyclePathway}--\eqref{eq:DirectPathway}. 
    }
    \label{fig:ScatteringMetaAtom}
\end{figure*}
\begin{figure*}[t]
    \includegraphics[width=\linewidth]{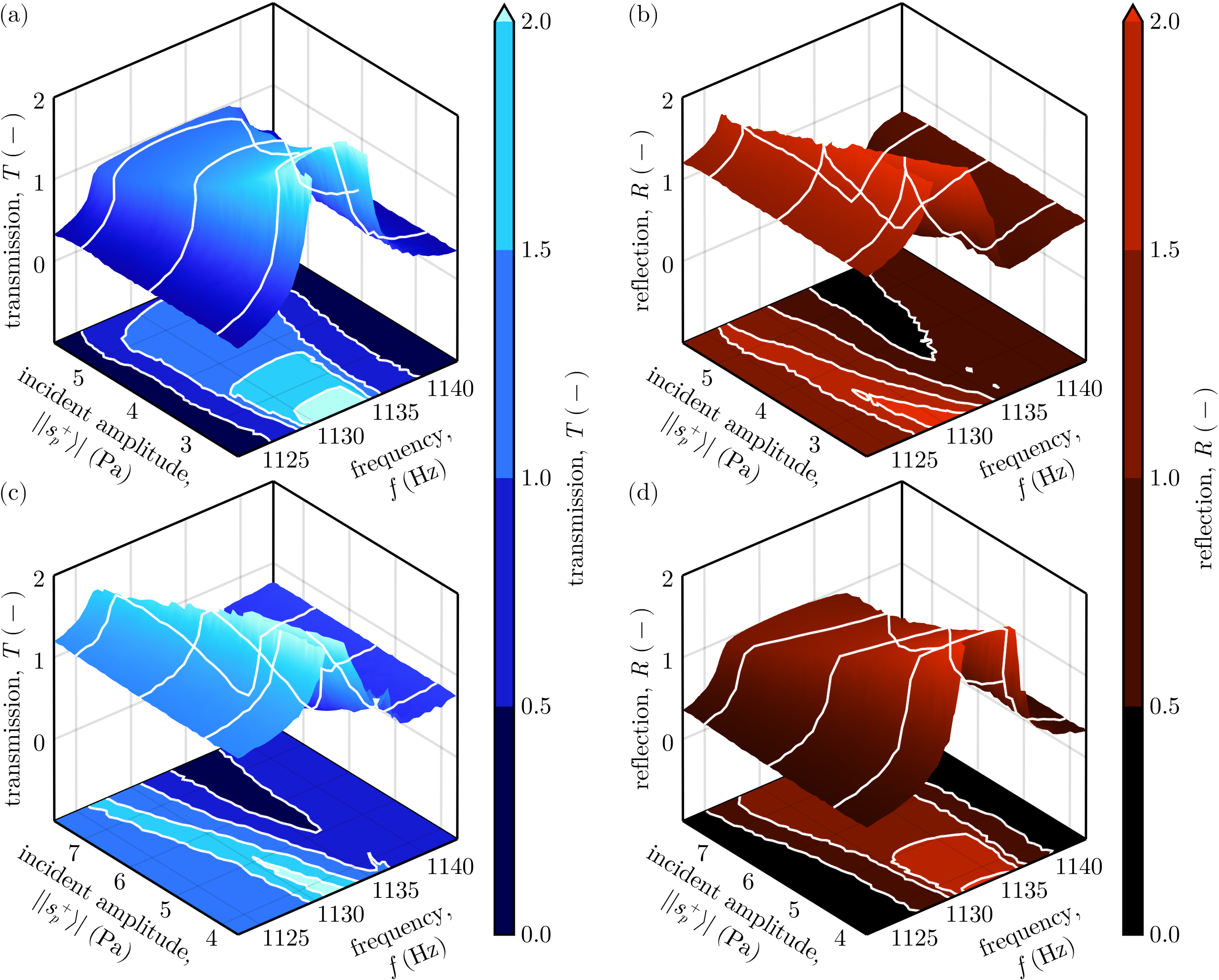}
    \caption{Experimental measurements of the transmission [left] and reflection [right] as function of incident frequency, $f$, and amplitude, $\left\vert\left\vert s_p^\mathrm{+}\right\rangle\right\vert$, for the direct [top] and side-branch configuration [bottom]. The full set of direct (a) transmission and (b) reflection data contains and complements the subsets presented in Fig.~\ref{fig:ScatteringMetaAtom}(b) [flow rate $\dot{Q}=9\,\mathrm{l/min}$]. Analogously, the side branch configuration's (c) transmission and (d) reflection extend the data discussed in Fig.~\ref{fig:ScatteringMetaAtom}(e) [flow rate $\dot{Q}=10\,\mathrm{l/min}$]. Note that Fig.~\ref{fig:ScatteringMetaAtom} shows measurement sets for fixed external excitation, however, the actual incident amplitude varies slightly with frequency due to small back-reflections from the terminations. In particular, the experimental data in Fig.~\ref{fig:ScatteringMetaAtom} is a flat projection of the white lines on the surfaces in (a)--(d).}
    \label{fig:AmplitudeSweep}
\end{figure*}
\begin{figure*}[t]
    \includegraphics[width=\linewidth]{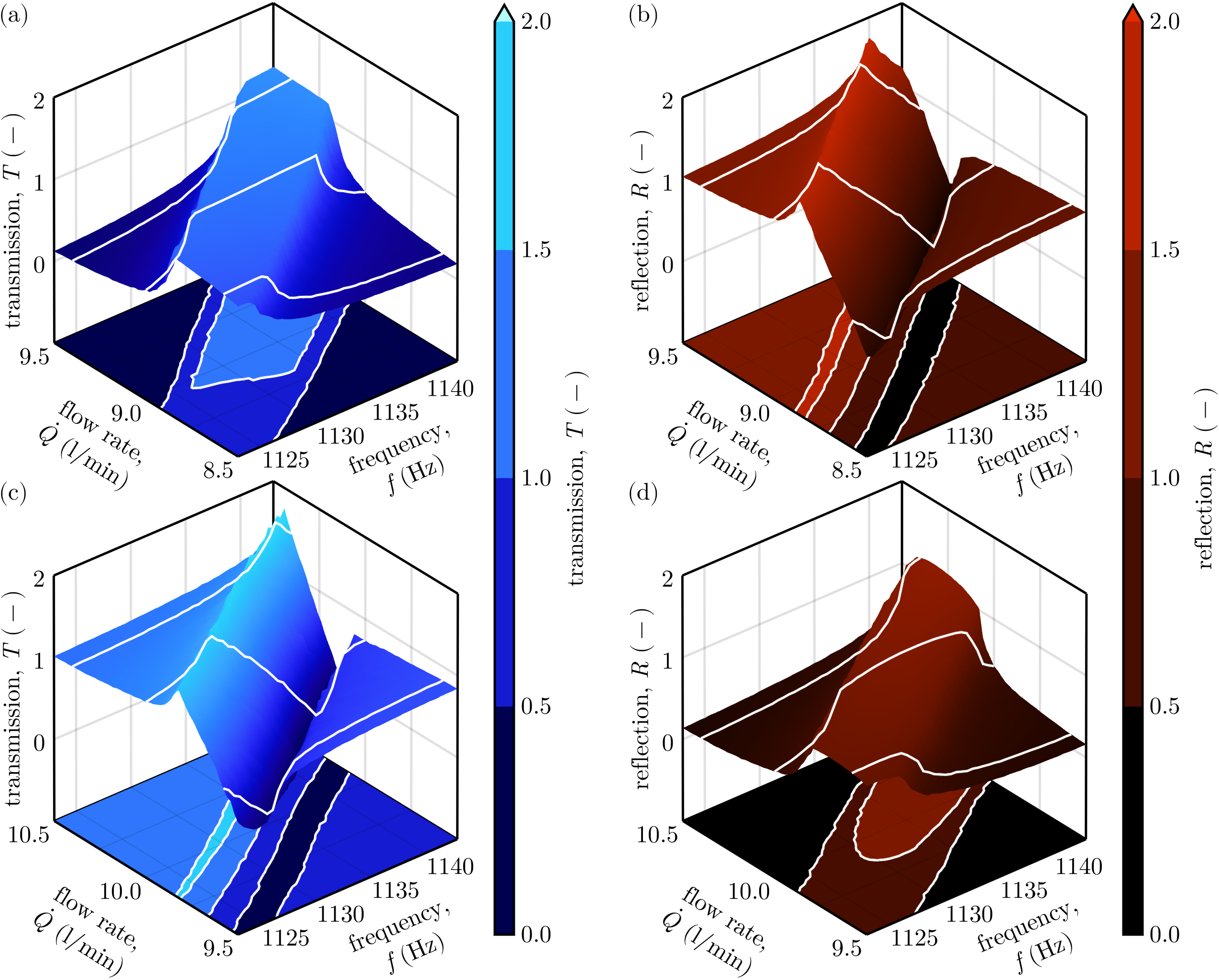}
    \caption{Experimental measurements of the transmission [left] and reflection [right] as function of incident frequency, $f$, and flow rate, $\dot{Q}$, for the direct [top] and side-branch configuration [bottom]. The full set of direct (a) transmission and (b) reflection data contains and complements the subsets presented in Fig.~\ref{fig:ScatteringMetaAtom}(c) [incident amplitude $\left\vert\left\vert s_p^\mathrm{+}\right\rangle\right\vert\approx5\,\mathrm{Pa}$]. Analogously, the side branch configuration's (c) transmission and (d) reflection extend the data discussed in Fig.~\ref{fig:ScatteringMetaAtom}(f) [incident amplitude $\left\vert\left\vert s_p^\mathrm{+}\right\rangle\right\vert\approx7.5\,\mathrm{Pa}$].}
    \label{fig:FlowSweep}
\end{figure*}
\subsection{Experimental measurements}\label{sec:ExperimentalMeasurements}
Fig.~\ref{fig:ScatteringMetaAtom}(a) presents the off-state [vanishing flow rate] measurements of the direct configuration [Fig.~\ref{fig:ExperimentalSetup}(a)], where the meta-atom acts as an acoustically rigid wall, exhibiting negligible transmission values [$\left\vert T\right\vert\approx0.05\approx\mathrm{const.}$] and (nearly) unitary reflection [$\left\vert R\right\vert\approx0.95\approx\mathrm{const.}$]. In contrast, for a nominal control signal [flow rate of $\dot{Q}=9~\mathrm{l/min}$] the device [in this configuration] exhibits self-sustained oscillations, which have profound effects on its behavior [Fig.~\ref{fig:ScatteringMetaAtom}(b)], resulting in transmission values on the order of unity [the transistor is activated]. Note that the scattering properties are dependent on the incident amplitude, an inherently nonlinear phenomenon, which leads to a perturbation filtering effect in the on-state. Specifically, higher intensity results in lower transmission [Fig.~\ref{fig:AmplitudeSweep}(a)], keeping the acoustic power downstream of the nonlinear scatterer quasi-constant, yet synchronous with the incident wave. Finally, Fig.~\ref{fig:ScatteringMetaAtom}(c) illustrates the dependence of the meta-atom's operational characteristics on the steady control signal [flow rate], for a fixed incident amplitude. Particularly, the frequency of self-oscillation, and by extension the location of the synchronization region in parameter space, is dependent on the DC input, allowing for some fine-tuning [Fig.~\ref{fig:FlowSweep}(a)].

Conversely, Fig.~\ref{fig:ScatteringMetaAtom}(d) presents the off-state measurements of the (conjugate) side-branch configuration [Fig.~\ref{fig:ExperimentalSetup}(b)], where the waveguide properties are mostly unaffected by the presence of the meta atom, resulting in negligible reflection values [$\left\vert T\right\vert\approx0.95\approx\mathrm{const.}$] and (nearly) unitary transmission [$\left\vert R\right\vert\approx0.05\approx\mathrm{const.}$]. Analogously to the case above, a nonzero control signal [flow rate of $\dot{Q}\approx10~\mathrm{l/min}$] and the corresponding limit cycle invert this behavior [Fig.~\ref{fig:ScatteringMetaAtom}(e)--(f)] and lead to amplitude [Fig.~\ref{fig:AmplitudeSweep}(b)] and flow rate [Fig.~\ref{fig:FlowSweep}(b)] dependent scattering. Note that the two configurations [Fig.~\ref{fig:ScatteringMetaAtom}(b)--(c) vs. Fig.~\ref{fig:ScatteringMetaAtom}(e)--(f)] exhibit complementary characters; the limit cycle contribution remains qualitatively similar, while the direct process [rigid wall vs. uniform waveguide] reverses the roles of transmission and reflection.

So far we have outlined the basic principle governing the proposed limit cycle scatterer, however, certain properties thereof warrant a detailed discussion. In particular, due to the synchronization based nature of the process, the bandwidth of the device is dependent on the Arnold tongue [cf.\ characteristic signature in Fig.~\ref{fig:AmplitudeSweep}]. Hence, it is subject to limitations imposed by the coupling interface and the incident acoustic waves. This has significant implications whenever frequency modulated signals need to be communicated across the meta-atom; higher saturable gain leads to (i) shorter relaxation times and, therefore, better signal quality, however, (ii) narrower synchronization tongue. Specifically, increasing the gain leads to a larger amplitude limit cycle, which is more resistant to perturbations, and thus, requires higher forcing intensity for the same synchronized bandwidth [Fig.~\ref{fig:AmplitudeSweep}]. Generally, frequency components outside the synchronization region are filtered out, resulting in the bandpass behavior seen in Fig.~\ref{fig:ScatteringMetaAtom}(b)--(c) [Fig.~\ref{fig:ScatteringMetaAtom}(e)--(f)]. However, should the entire spectrum of the incident wave be outside the Arnold tongue [no synchronization occurs], the output signal will contain significant energy at the self-oscillation frequency of the nonlinear scatterer [cf.\ App.~\ref{app:SynchronizationSpectrum}]. Furthermore, additional considerations may be needed for dynamical modulation close to the synchronization boundary \cite{Lucas2018}. Finally, note that all aforementioned effects are the product of a weakly nonlinear system [cf.\ App.~\ref{app:HigherHarmonics}], hence, signal distortions caused by exciting harmonics of the fundamental frequency are negligible.

\subsection{Theoretical model}\label{sec:TheoreticalModel}
Globally stable shear layers can amplify acoustic perturbations at certain frequencies \cite{Bourquard2021}, a phenomenon driving the self-sustained oscillations in our meta-atom. In particular, a steady grazing flow over an aperture exhibits a shear layer along which vorticity perturbations are advected, which leads to energy transfer between the mean flow and the sound field. The resulting energy balance is governed by the time averaged projection of the coherent fluctuations of the Lamb vector field onto the acoustic field. This mechanism can give rise to superradiant scattering of sound waves in a specific frequency band and, under suitable geometric conditions, i.e., coalescence of the Helmholtz mode with this amplifying range, to self‑oscillation. Nevertheless, this response eventually saturates due to nonlinear effects \cite{Bourquard2021,Aperture}, yielding a finite amplitude limit cycle governed by the balance between saturable gain and radiation (and viscous) losses. Furthermore, coupling to the waveguide allows interactions with the incident waves, which can perturb the oscillator, resulting in a nonlinear scattering process.

In the following we model this behavior with a single variable [cf. App.~\ref{app:HigherHarmonics}] Liénard system of the form
\begin{equation}\label{eq:OscillatorEquation}
    \ddot{p}+\mu\left[1 + \psi(p)\right]\dot{p}+p = \left\langle\kappa\big\vert s^{\mathrm{+}}\right\rangle\,,
\end{equation}
where $p$ denotes the acoustic pressure inside the cavity, normalized by its unperturbed amplitude, $\left\vert p_p\right\vert$, $\mu$ incorporates the radiation and viscous losses and $\left\langle\kappa\right\vert=\kappa\left\langle1\right\vert+\kappa\left\langle2\right\vert$ stands for the direct coupling matrix. Conversely, $\left\vert s^{\mathrm{+}}\right\rangle = s\left\vert1\{2\}\right\rangle\cos{\Omega\tau}$ represents the incident wave, with $s=\left\vert\left\vert s_p^{\mathrm{+}}\right\rangle\right\vert/\left\vert p_p\right\vert$ being the amplitude of the physical wave, $\left\vert\left\vert s_p^{\mathrm{+}}\right\rangle\right\vert$, normalized by $\left\vert p_p\right\vert$ and $\Omega=f/f_0$ denoting the ratio between incident wave and self-oscillation frequencies. Finally, $\psi(p)$ indicates the nonlinearity
\begin{equation}\label{eq:NonlinearLaw}
    \psi(p) =  \frac{3p\sinh^{-1}{p}}{(1+p^2)^{5/2}} - \frac{2-p^2}{(1+p^2)^2}\,.
\end{equation}
For $\mu>0$ Eq.~\eqref{eq:OscillatorEquation} has a unique, stable and globally attractive limit cycle, which is topologically similar to the trajectories of the conservative system \cite[Corollary 1]{Synchronization}. Hence, under the assumption of a small [cf.\ App.~\ref{app:HigherHarmonics}] bifurcation parameter, $\mu\ll1$, we invoke Bogoliubov's theorem \cite{Krylov1950} and separate the slow timescale evolution of amplitude and phase from the carrier signal, obtaining
\begin{equation}\label{eq:LimitCyclePathway}
  \frac{\mathrm{d}a}{\mathrm{d}\tau} = \left(j\frac{1-\Omega^2}{2\Omega}+\frac{\mu}{2}\frac{1-\vert a\vert^2}{1+\vert a\vert^2}\right)a - \frac{j}{2\Omega}\left\langle\kappa\big\vert s^{\mathrm{+}}\right\rangle\,,
\end{equation}
where $\mathrm{d}/\mathrm{d}\tau$ represents derivative w.r.t.\ time, normalized by the self-oscillation frequency, $f_0$, and $j$ denotes the imaginary unit. Conversely, $a$ is the complex envelope of the normalized pressure limit cycle inside the whistle, with $\vert a\vert$ and $\arg\{a\}$ describing the slow time scale magnitude and phase behavior, respectively.

To study the scattering w.r.t.\ an external excitation, we need to consider two distinct processes \cite{Fan2003}; (i) the direct pathway bypassing the meta-atom's dynamics and (ii) radiation by the perturbed limit cycle
\begin{equation}\label{eq:DirectPathway}
    \left\vert s^{\mathrm{-}}\right\rangle = \mathbf{S}\left\vert s^{\mathrm{+}}\right\rangle + \left\vert \tilde\kappa\right\rangle a\,,
\end{equation}
where the complex valued $\mathbf{S}\approx\mathbf{1}$ represents the scattering matrix of the direct process and $\left\langle\tilde\kappa\right\vert=\tilde\kappa\left\langle1\right\vert+\tilde\kappa\left\langle2\right\vert$ denotes the conjugate coupling matrix. Note that Eqs.~\eqref{eq:LimitCyclePathway}--\eqref{eq:DirectPathway} are (i) evolving in a time dependent reference frame, (ii) containing terms deviating from classical temporal coupled mode theory (TCMT) \cite{Fan2003} and the nonlinear extension assumed in \cite{Pedergnana2023,Pedergnana2025}, and (iii) breaking Lorentz reciprocity \cite{Zhao2019}. Specifically, (i) we perform asymptotic averaging in a coordinate system moving in-phase with the incident wave, i.e., $\left\vert s^{\mathrm{+}}\right\rangle=\left\vert1\{2\}\right\rangle s$ with $s=\mathrm{const}$, (ii) which introduces an asymmetric detuning term, $j(1-\Omega^2)/2\Omega$, and multiplicative forcing factor, $j/2\Omega$. Furthermore, (iii) we assume the coupling terms between incident wave and self-oscillator amplitude are nonreciprocal, i.e., the coupling strength depends on the direction of the energy flow [$\kappa\neq\tilde\kappa$]. Those modifications are necessary to qualitatively (and quantitatively) capture the physical processes involved (cf.\ App.~\ref{app:ModelCoefficients}).

From Eqs.~\eqref{eq:LimitCyclePathway}--\eqref{eq:DirectPathway} we can also derive closed form expressions for the reflection and transmission coefficients in the synchronized steady state
\begin{align}
    R &= S_{11}+\frac{\tilde\kappa\kappa}{2\Omega}\left(\frac{1-\Omega^2}{2\Omega}-j\frac{\mu}{2}\frac{1-\vert a\vert^2}{1+\vert a\vert^2}\right)^{-1}\,, \label{eq:Reflection} \\
    T &= S_{21}+\frac{\tilde\kappa\kappa}{2\Omega}\left(\frac{1-\Omega^2}{2\Omega}-j\frac{\mu}{2}\frac{1-\vert a\vert^2}{1+\vert a\vert^2}\right)^{-1}\,, \label{eq:Transmission}
\end{align}
with the solutions for $\vert a\vert$ and the specific parameter values discussed in App.~\ref{app:ModelCoefficients}. 

Comparison between experimental and theoretical results [Fig.~\ref{fig:ScatteringMetaAtom}] shows good agreement, confirming the chosen saturable nonlinearity and the validity of Eqs.~\eqref{eq:OscillatorEquation}--\eqref{eq:Transmission}. Note that the TCMT Eqs.~\eqref{eq:LimitCyclePathway}--\eqref{eq:DirectPathway} describe the system accurately throughout the considered subset of parameter space [cf.\ Fig.~\ref{fig:ScatteringMetaAtom}], including outside the synchronization region, where additional frequency components are generated. However, Eqs.~\eqref{eq:Reflection}--\eqref{eq:Transmission} are valid only for synchronized conditions [the usual state of interest], since they rely on a steady state [$a=\mathrm{const.},~\mathrm{d}a/\mathrm{d}\tau=0$], which is only the case within the Arnold tongue.

\section{Conclusion}\label{sec:Conclusion}
In this theoretical and experimental study we demonstrate a self-oscillating meta-atom, implemented as a flow-driven acoustic oscillator and functioning as an acoustic transistor. By leveraging a steady external flow as a control signal, the device transitions between a fully reflective off-state and a transmissive on-state, where synchronization between the self-sustained limit cycle and incident acoustic waves governs the energy transfer. Experimental measurements reveal that the transmission bandwidth is dictated by the geometry of the Arnold tongue in parameter space, while the nonlinear dependence on incident amplitude enables perturbation filtering, stabilizing downstream acoustic power. Theoretical modeling based on a Liénard-type oscillator featuring saturable gain and a linear loss quantitatively captures these effects, validating the interplay between synchronization, nonlinearity, and scattering.

This work highlights the potential of integrating self-oscillation into acoustic metamaterials. Apart from conventional approaches relying on static nonlinearity or topological states, synchronization-driven interactions may offer a dynamic, field-independent mechanism for wave control, reconciling concepts from complex systems with metamaterial design. The application possibilities go beyond the demonstrated transistor-like behavior; e.g., adaptive amplitude filtering via nonlinearity, or nonreciprocal signal routing via the synchronization topology of a network of self-oscillators \cite{LimitCycleDiode}. Indeed, future work could explore cascaded synchronization networks leveraging collective oscillator dynamics or multi-frequency operation, to name few, establishing a new venue of metamaterials research.

\appendix

\section{Model coefficients}\label{app:ModelCoefficients}
\begin{figure}
    \includegraphics[width=\linewidth]{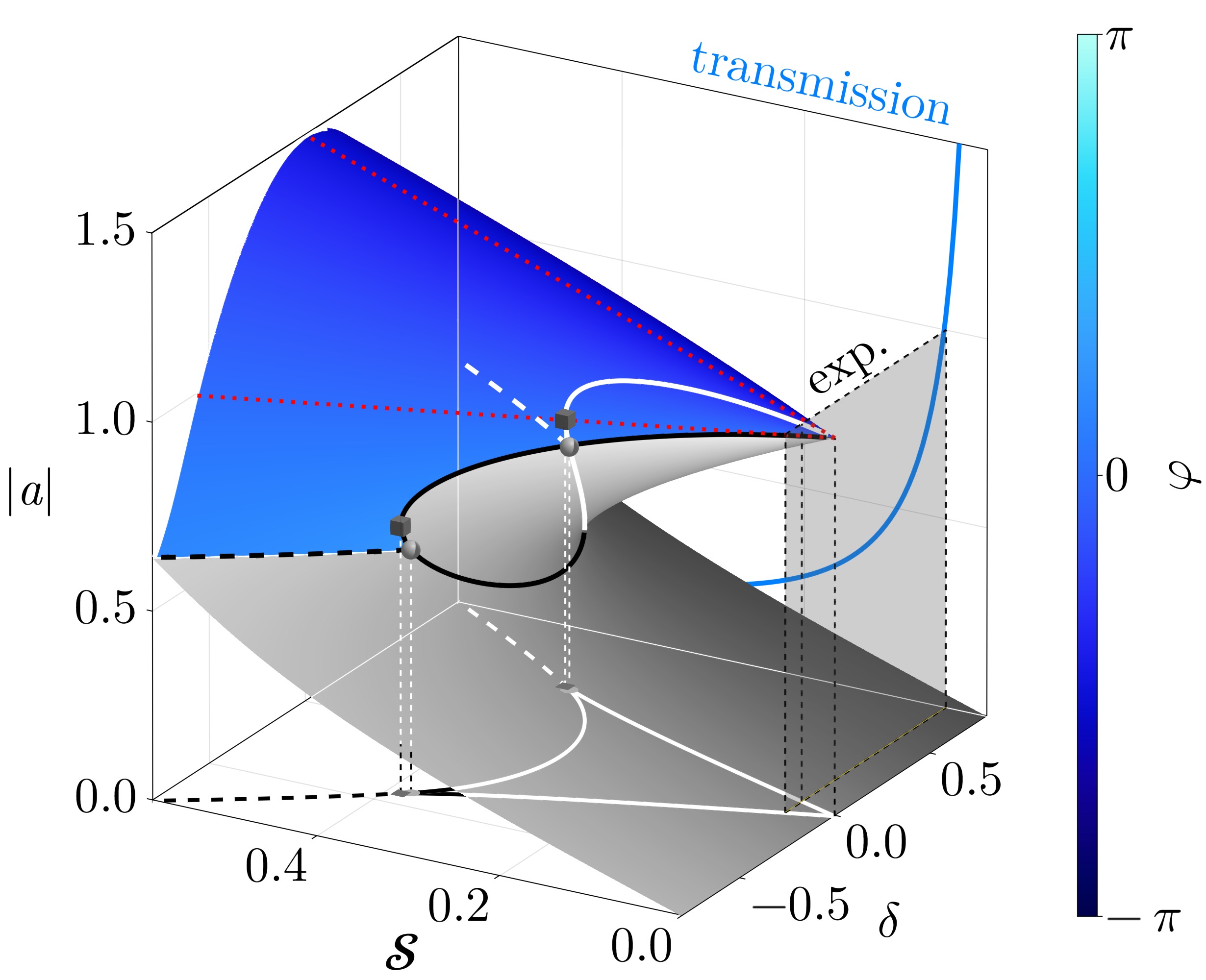}
    \caption{Steady state (synchronized) solutions of Eq.~\ref{eq:LimitCyclePathway}; amplitude, $\vert a\vert$, phase, $\varphi$, and stability, as a function of the two dimensional parameter space $\mathcal{S}-\delta$. The phases of stable solutions are depicted by the color map, while unstable ones are shaded gray. Solid lines indicate the continuous branches of fold (of cycles) bifurcations and their projection onto the $\mathcal{S}-\delta$ plane. Dashed lines depict branches of Neimark-Sacker bifurcations and their projection. Cube markers with square projection represent cusp (of cycles). Sphere markers with circular projection indicate Bogdanov-Takens (eigenvalue-vector degeneracy) bifurcations. Red lines indicate the $\vert a\vert=1$ [left] and $\delta=0$ [top] level sets. The second $z$--axis (light blue) shows the transmission (blue curve) at $\delta=0$.}
    \label{fig:Theory}
\end{figure}
\begin{table}[b]
    \vspace{-5mm}
    \centering
    \caption{Measured model parameters.}
    \label{tab:MeasuredCoefficients}
    \setlength{\tabcolsep}{9pt}
    \begin{tabular}{cccc}
        \toprule
        configuration & $\left\vert p_p\right\vert~(\mathrm{Pa})$ & $S_{11}~(\mathrm{-})$ & $S_{12}~(\mathrm{-})$ \\
        \midrule
        direct & $185$ & $0.935$ & $0.070\cdot\mathrm{e}^{-2.74j}$ \\
        side-branch & $270$ & $0.050$ & $0.965\cdot\mathrm{e}^{1.90j}$ \\
        \bottomrule
    \end{tabular}
\end{table}
In the steady state [$\mathrm{d}a/\mathrm{d}\tau=0$] Eq.~\eqref{eq:LimitCyclePathway} yields
\begin{equation}
    \left(1-\vert a\vert^2\right)^2\vert a\vert^2+\delta^2(1+\vert a\vert^2)^2\vert a\vert^2=(1+\vert a\vert^2)^2\mathcal{S}^2\,,
\end{equation}
with $\delta=2\Delta/\mu$, $\Delta=(1-\Omega^2)/2\Omega$ and $\mathcal{S}=\left\vert\left\langle\kappa\big\vert s^{\mathrm{+}}\right\rangle\right\vert/\mu\Omega$. This reduces to a cubic equation in $\vert a\vert^2$
\begin{equation}\label{eq:CubicA}
    c_1\vert a\vert^6+c_2\vert a\vert^4+c_3\vert a\vert^2+c_4=0\,,
\end{equation}
with coefficients
\begin{align*}
    c_1 &= 1+\delta^2\,, &&c_2 = -2+2\delta^2-\mathcal{S}^2\,, \\
    c_3 &= 1+\delta^2-2\mathcal{S}^2\,, &&c_4 = -\mathcal{S}^2\,. \\
\end{align*}
Eq.~\eqref{eq:CubicA} has up to three solutions [Fig.~\ref{fig:Theory}] with the (unconditionally) stable one given by
\begin{equation}
    \vert a\vert^2 = \frac{2\sqrt{r}}{3c_1}\cos\left(\frac{1}{3}\cos^{-1}\left(\frac{q}{2p^{\frac{3}{2}}}\right)\right)-\frac{c_2}{3c_1}\,,
\end{equation}
where
\begin{align*}
    r = c_2^2-3c_1c_3\,, \quad q = 2c_2^3-9c_1c_2c_3+27c_1^2c_4\,.
\end{align*}

Next, we discuss independent magnitude estimates of the coefficients used in Sec.~\ref{sec:Results}. We focus on the direct configuration, where the side branch follows analogously.
\begin{enumerate}
    \item The amplitude of the unperturbed limit cycle can be estimated from the probability density function of the cavity pressure, $\left\vert p_p\right\vert\approx185\,\mathrm{Pa}$, and the eigenfrequency -- from its spectrum, $f_0\approx1133\,\mathrm{Hz}$.
    \item In the absence of an incident wave and under a nominal control signal [flow rate of $\dot{Q}=9~\mathrm{l/min}$] the meta-atom (i) exhibits an unperturbed pressure limit cycle with an amplitude of $\left\vert p_p\right\vert\approx185\,\mathrm{Pa}$ and (ii) radiates acoustic waves in the waveguide with an amplitude of $\left\vert\left\vert s_p^\mathrm{-}\right\rangle\right\vert\approx5\,\mathrm{Pa}$. Hence, we can estimate the conjugate coupling coefficient as
    \begin{equation}
        \left\vert\tilde\kappa\right\vert\approx\frac{\left\vert\left\vert s_p^\mathrm{-}\right\rangle\right\vert}{\left\vert p_p\right\vert}=\frac{5\,\mathrm{Pa}}{185\,\mathrm{Pa}}\approx0.027.
    \end{equation}
    \item The steady state solutions of Eq.~\ref{eq:LimitCyclePathway} [Fig.~\ref{fig:Theory}] permit a frequency for any incident wave intensity, such that the resulting limit cycle amplitude is identical to the unperturbed one. This allows to eliminate the term containing the bifurcation parameter, $\mu$, and to determine the direct coupling coefficient. In particular, for an incident amplitude of $\left\vert\left\vert s_p^\mathrm{+}\right\rangle\right\vert\approx5\,\mathrm{Pa}$ at a frequency of $f=1127\,\mathrm{Hz}$ we have
    \begin{equation}
        \left\vert\kappa\right\vert\approx\left\vert\frac{1-\Omega^2}{s}\right\vert=\frac{1-\left(\frac{1127\,\mathrm{Hz}}{1133\,\mathrm{Hz}}\right)^2}{\frac{5\,\mathrm{Pa}}{185\,\mathrm{Pa}}}\approx0.39.
    \end{equation}
    \item Conversely, if the incident frequency is identical to the eigenfrequency, the bifurcation parameter, $\mu$, can be estimated based on $\left\vert\kappa\right\vert$. For an incident amplitude of $\left\vert\left\vert s_p^\mathrm{+}\right\rangle\right\vert\approx5\,\mathrm{Pa}$ at a frequency of $f=1133\,\mathrm{Hz}$ we have $\vert a\vert\approx1.07$ and consequently
    \begin{equation}
        \mu\approx2\left\vert\frac{\left\vert\kappa\right\vert s}{\vert a\vert}\frac{1+\vert a\vert^2}{1-\vert a\vert^2}\right\vert\approx0.28.
    \end{equation}
\end{enumerate}
Note, that some of the coefficients above are directly adopted in the calculations, Tab.~\ref{tab:MeasuredCoefficients}, and some are further refined by fitting to the scattering measurements,~Tab.~\ref{tab:FittedCoefficients}.

Finally, note the Bogdanov-Takens bifurcation point exhibits an eigenvalue-vector degeneracy, and consequently, a chiral behavior similar to an exceptional point.
\begin{table}[b]
    \vspace{-5mm}
    \centering
    \caption{Fitted model parameters.}
    \label{tab:FittedCoefficients}
    \setlength{\tabcolsep}{6pt}
    \begin{tabular}{ccccc}
        \toprule
        conf. & $f_0~(\mathrm{Hz})$ & $\mu~(\mathrm{-})$ & $\tilde\kappa~(\mathrm{-})$ & $\kappa~(\mathrm{-})$ \\
        \midrule
        dir. & $1132.2$ & $0.233$ & $0.027\cdot\mathrm{e}^{1.233j}$ & $0.389\cdot\mathrm{e}^{1.233j}$ \\
        s-b  & $1132.5$ & $0.067$ & $0.025\cdot\mathrm{e}^{2.116j}$ & $0.376\cdot\mathrm{e}^{2.116j}$ \\
        \bottomrule
    \end{tabular}
\end{table}
\newpage
\section{Synchronization spectrum}\label{app:SynchronizationSpectrum}
\begin{figure}[t]
    \includegraphics[width=\linewidth]{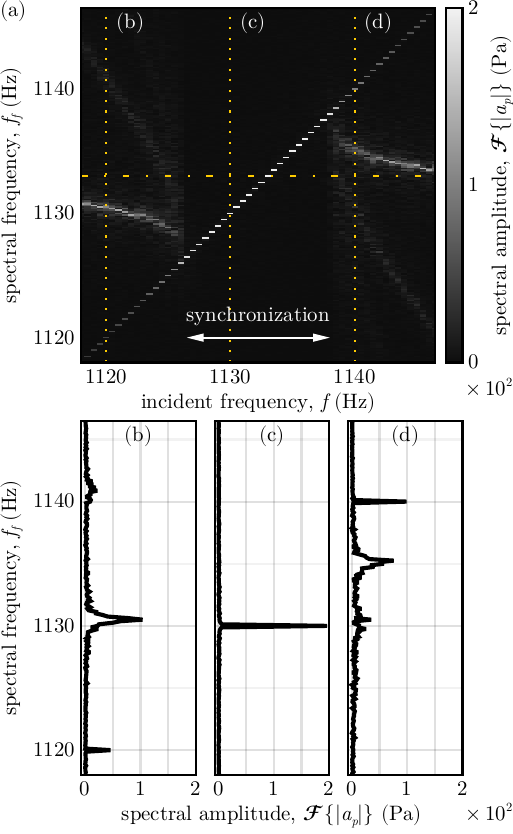}
    \caption{Spectrum of the self-oscillator for a synchronization process under an incident wave amplitude of $\left\vert\left\vert s_p^\mathrm{+}\right\rangle\right\vert\approx5\,\mathrm{Pa}$ [direct configuration]. Panel (a) shows the Fourier components of the (physical) oscillator pressure, $p_p$, as function of the incident frequency, $f$, and Fourier-frequency, $f_f$. The energy can be contained mainly at a single frequency or spread over multiple ones. In (b), (c), and (d) three distinct cases are shown, corresponding to incident frequencies indicated by the vertical dotted lines in (a), respectively. The horizontal dash-dotted line indicates the self-oscillation frequency. For an incident frequency (b) away from the synchronization region ($f=1120\,\mathrm{Hz}$]) the main peak corresponds to the self-oscillation frequency [$\approx1133\,\mathrm{Hz}$], when (c) inside the region the signal is mainly monochromatic, and (d) in the vicinity thereof the incident frequency dominates.
    }
    \label{fig:Spectrum}
\end{figure}
Figure \ref{fig:Spectrum}(a) presents a two-dimensional plot where the horizontal axis represents the incident frequency $f$, and the vertical axis represents the Fourier frequency $f_f$. The color (intensity) of the plot corresponds to the magnitude of the Fourier components of the oscillator's (physical) pressure, $|a_p|$. The spectrum reveals distinct behaviors depending on the ratio between incident and natural frequency of the oscillator, $\Omega=f/f_0$. Specifically, three key regions are highlighted in panels (b), (c), and (d).

Panel (b) corresponds to incident frequencies far from the synchronization region. Here, the spectrum shows a dominant peak close to the self-oscillation frequency and a lesser one at the incident frequency. This indicates that the oscillator primarily oscillates at its natural frequency, with weak influence from the external forcing.

Panel (c) illustrates the spectrum within the synchronization region. In this case, the oscillator phase-locks onto the incident wave, resulting in a nearly monochromatic signal with a single peak at the incident frequency. This demonstrates the synchronization effect, where the oscillator's dynamics are entrained by the external drive.

Panel (d) shows the spectrum in the vicinity of the synchronization region. In this transitional area we observe a dominant peak at the incident frequency, with additional ones present between it and the self-oscillation frequency, indicating proximity to a co-dimension 1 bifurcation line \cite[pp. 58--61]{Balanov2008}, a fold of cycles in this case [cf.\ Fig.~\ref{fig:Theory}].

\section{Higher harmonics}\label{app:HigherHarmonics}
\begin{figure}[b]
    \includegraphics[width=\linewidth]{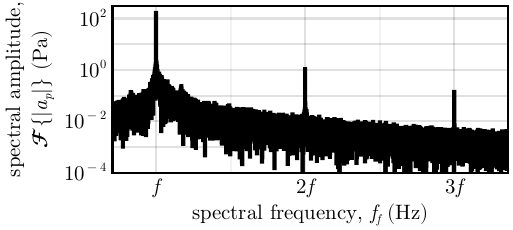}
     \caption{Comparison between fundamental frequency and higher harmonics in the spectrum of the self-oscillator for $f=1130\,\mathrm{Hz}$. Note that the component at the first harmonic is more than two orders of magnitude lower than the one at the fundamental frequency. This indicates that our system is a weakly nonlinear self-oscillator, which justifies the averaging procedure in Sec.~\ref{sec:TheoreticalModel}. Furthermore, the absence of any discernible peaks at frequencies incommensurate with the fundamental indicates that only the uniform acoustic pressure (Helmholtz) mode constructively interacts with the shear layer as an aeroacoustic limit cycle, which validates modeling the system with a single dependent variable, $a(t)\in\mathbb{C}$. 
     }
    \label{fig:Harmonics}
\end{figure}
Figure \ref{fig:Harmonics} presents a detailed comparison between fundamental frequency and higher harmonics in the spectrum of the self-oscillator. The analysis provides compelling evidence of the system's weak nonlinearity and the dominance of the fundamental mode. These observations validate the use of averaging techniques and the single-mode approximation in our theoretical model.

\begin{SCfigure*}
    \includegraphics[width=375pt]{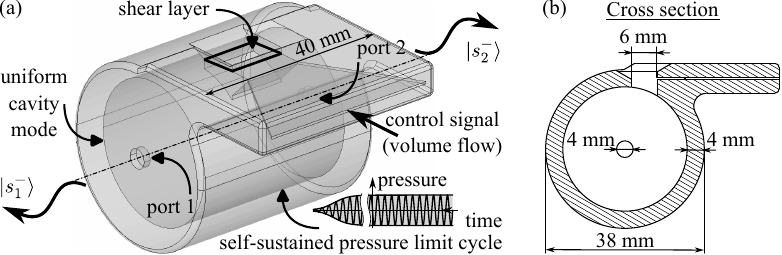}
    \caption{Geometry of the self-oscillating meta-atom, with (a) three dimensional isometric view and (b) cross section along the air supply line. The main properties of the system are schematically depicted in (a) while the specific dimensions are given in (b).}
    \label{fig:WhistleGeometry}
\end{SCfigure*}
\section{Meta-atom geometry}\label{app:WhistleGeometry}
Figure~\ref{fig:WhistleGeometry} illustrates the geometry of the meta-atom designed for this study. Specifically, sub-panel (a) shows a semi-transparent isometric view emphasizing the spatial pressure distribution of the self-oscillating mode and the main components of the system, such as, the cavity, the coupling ports, the air supply line, and the shear layer. Moreover, sub-panel (b) depicts all relevant dimensions and the internal structure of the cavity.
\bibliography{references}

\end{document}